\documentstyle[12pt]{article}
\input psfig
\newcommand{\la}[1]{\label{#1}}

\newlength{\numlen}

\newlength{\indexlength}

\newcommand{\be}{\begin{equation}}
\newcommand{\ee}{\end{equation}}
\newcommand{\ba}{\begin{eqnarray}}
\newcommand{\ea}{\end{eqnarray}}

\newcommand{\rmi}[1]{{\mbox{\scriptsize #1}}}

\newcommand{\tr}{\mbox{Tr\,}}

\newcommand{\bfx}{\mbox{\bf x}}

\newcommand{\fr}[2]{{\frac{#1}{#2}}}

\begin{document}

\begin{titlepage}
\begin{flushright}
CERN-TH.7244/94\\
IUHET-279\\
\end{flushright}
\begin{centering}
\vfill

{\bf THE ELECTROWEAK PHASE TRANSITION AT {\Large $m_H \simeq m_W$}}
\vspace{1cm}

 K. Farakos$^a$\footnote{Partially supported by a CEC Science program
(SCI-CT91-0729).}, K. Kajantie$^{b}$, K. Rummukainen$^c$ and M.
Shaposhnikov$^d$\footnote{On leave of absence from the Institute for
Nuclear Research of the Russian Academy of Sciences, Moscow 117312,
Russia.} \\

\vspace{1cm}
{\em $^a$National Technical University of Athens, Physics
Department,\\ Zografou Campus, GR 157 80, Athens, Greece\\}
\vspace{0.3cm}
{\em $^b$Department of Theoretical Physics,
P.O. Box 9, 00014 University of Helsinki, Finland\\}
\vspace{0.3cm}
{\em $^c$Indiana University, Department of Physics,
Swain Hall West 117,\\Bloomington IN 47405, USA\\}
\vspace{0.3cm}
{\em $^d$Theory Division, CERN,\\ CH-1211 Geneva 23, Switzerland}

\vspace{2cm}
{\bf Abstract}

\end{centering}

\vspace{0.3cm}\noindent
We study the finite temperature electroweak transition with
non-perturbative lattice Monte Carlo simulations. We find that
it is of first order, at least for
Higgs masses up to 80 GeV. The critical temperature of the
phase transition is found to be smaller than that determined by a
2-loop renormalization group improved effective potential. The jump
of the order parameter at the critical temperature is considerably
larger than the perturbative value. By comparing lattice
data and perturbation theory, we demonstrate that the latter, for the
computation of the vacuum expectation value of the
Higgs field $v(T)$ in the broken phase at given temperature,
converges quite well, provided $v(T)/T>1$. An upper bound on the
Higgs
mass necessary for electroweak baryogenesis in the light of the
lattice data is briefly discussed.

\vskip 0.5in
\noindent CERN-TH.7244/94\\
\noindent May 1994
\addtocounter{page}{1}
\thispagestyle{empty}
\pagebreak

\end{titlepage}
\section{Introduction}
The high-temperature limit of the 4-dimensional electroweak theory
near
the phase transition and for sufficiently large Higgs masses can be
described by an effective 3-dimensional gauge-Higgs theory (for a
discussion of high-temperature phase transitions see \cite{kir}, for
a general consideration of dimensional reduction see \cite{3d4d},
for 3d EW theory see \cite{fkrs,kari,jakovac}). The effective
theory is strongly coupled in the symmetric phase \cite{linde,gpy};
in
particular, pure SU(2) gauge theory in 3d is confining. This fact
makes a reliable {\em entirely perturbative} computation of almost
all
characteristics of the phase transition impossible. For example, the
computation of the critical temperature $T_c$ requires the comparison
of the vacuum energy of the symmetric and broken phases, and
the former cannot be estimated on perturbative grounds. The bubble
nucleation temperature $T^*$ in cosmology is related to the structure
of the effective action for small scalar fields, which is
precisely the place
where infrared divergences are most severe. The expectation
value of the Higgs field $v(T^*)$ at the temperature $T^*$, necessary
for the estimate of the fermion number non-conservation rate
\cite{krs}, cannot be found perturbatively, just because $T^*$ is
non-perturbative. At the same time, the
determination of the vacuum expectation value of the Higgs field in
the broken phase at a {\em given} temperature $T$ by perturbation
theory does not require any information on the symmetric phase and
may
be good enough provided that the ratio $v(T)/T$ is sufficiently
large.
Therefore, the study of the nature of the electroweak phase
transition requires non-perturbative methods. The most
straightforward ones are lattice Monte--Carlo simulations.

First lattice Monte Carlo results on the electroweak phase transition
using a 3d effective theory have already been given in \cite{kari}
(for 4d simulations see \cite{bunk}). We improve here the analysis of
\cite{kari} in two ways: by going from the 1-loop to the 2-loop level
in the discussion of the 3d effective theory and its effective
potential, and by extending considerably the numerical calculations.
We find that the EW phase transition is of the first order at $m_H
\simeq m_W$. The critical temperature we determined is smaller than
that following from the perturbative analysis, while the jump of the
order parameter is larger. At the same time, 2-loop effective
potential gives an adequate description of the order parameter in the
{\em broken} phase for $v(T)/T>1$. These results are in a qualitative
agreement with a picture of the electroweak phase transition
suggested in \cite{ms}. Given the information provided by lattice
simulations we comment shortly on the question of the Higgs mass
necessary for electroweak baryogenesis \cite{s:higgs,ms}.

The advantages of considering the effective
3d vacuum gauge-Higgs system in comparison with the full
4d theory at non-zero temperatures have already been
discussed in \cite{kari} (from the lattice point of view) and in
\cite{fkrs} (from the perturbation theory point of view).
In the present paper, for reasons of simplicity, the U(1) factor
of the complete electroweak theory is disregarded. The fermionic
contributions are omitted as well\footnote{The most important
fermionic contribution is that of the top quark to the effective
scalar mass at high temperatures. It does change the absolute value
of the critical temperature, decreasing it, but changes only
marginally
dimensionless ratios such as $v(T_c)/T_c$, most important for
cosmological applications.}.

The paper is organized as follows. In Section 2 we briefly
summarize  the results from the perturbative analysis.
The lattice action is given in Section 3, as well as a
discussion on relating lattice numbers to physical quantities.
Numerical results are presented in the Section 4. Section 5 is
discussion.
\section{Continuum Lagrangian}
The continuum Lagrangian of the effective 3d theory under
consideration is
\begin{eqnarray}
&&S_{\rmi{eff}}
= \int d^3x \biggl\{{1\over4} F_{ij}^aF_{ij}^a +
\fr12 (D_iA_0)^a(D_iA_0)^a + (D_i\phi)^\dagger(D_i\phi)+
\nonumber
\\&&
\fr12 m_D^2 A_0^aA_0^a + \frac{1}{4}\lambda_A (A_0^aA_0^a)^2
+ m_3^2\phi^\dagger\phi
+\lambda_3(\phi^\dagger\phi)^2
+ h_3 A_0^aA_0^a \phi^\dagger\phi \biggr\}.
\label{3daction}
\end{eqnarray}
It contains gauge fields together with the Higgs doublet and a scalar
triplet (the former time component of the gauge field).
Here all bosonic fields have the canonical dimension [GeV]$^{1\over
2}$ and 3d gauge and scalar couplings  $g_3^2,~ \lambda_3,~\lambda_A$
and
$h_3$ have dimension [GeV]. The relation between 4d and 3d coupling
constants and masses on the 1- and 2-loop levels has been  discussed
in
\cite{fkrs}. For all numerical and analytical work we use the
following simplified relations:
\be
g=\fr23,\qquad g_3^2 = g^2 T,\qquad\lambda_3 =
\frac{1}{8}g_3^2\frac{m_H^2}{m_W^2},\qquad
h_3 = \frac{1}{4}g_3^2, \qquad\lambda_A = 0.
\ee
Due to the super-renormalizability of the 3d theory (only a finite
number of diagrams are divergent) these relations are
scale-independent\footnote{The actual value of the coupling
$\lambda_A$ is rather small, $17 g^4 T/(48 \pi^2)$. We included
this coupling in the lattice simulations just to make sure that the
lattice action is bounded from below for constant fields, and
checked that the results are independent of $\lambda_A$ provided
that it is small enough.}. The Debye screening mass and the
effective Higgs mass are
\cite{fkrs}
\be
m_D^2 = \frac{5}{6} g^2 T^2,
\label{mD}
\ee
\be
m_3^2(\mu_3) = \biggl[{3\over16} g_3^2 T + \fr12 \lambda_3 T +
\frac{g_3^2}{(4
\pi)^2}\biggl(\frac{149}{96}g_3^2 +
\fr34 \lambda_3\biggr)\biggr] - \fr12
m_H^2+{1\over16\pi^2}\biggl[
f_{2m}\biggl(\log{3T\over\mu_3}+c\biggr)\biggr].
\ee
The numerical constant $c=-0.348725$, the parameter $\mu_3$ is the
scale
of the ${\overline{{\rm MS}}}$ scheme. The Debye screeening mass
$m_D$ is
a 3d renormalization group invariant, but the effective Higgs mass
runs
with the normalization point due to logarithmic divergences on the
2-loop level. The 2-loop coefficient $f_{2m}$ is given by
\be
f_{2m} = {81\over16}g_3^4+9\lambda_3 g_3^2-12\lambda_3^2.
\label{count}
\ee

The 2-loop renormalization group improved effective potential for
this theory has been computed in \cite{fkrs}. (The computation of the
high-temperature limit of the 2-loop effective potential in 4d has
been
performed in \cite{ae}). We do not write the expression here for
lack of space and just mention that, for $m_H = 80$ GeV and $m_W =
80.6$ GeV, the critical temperature and the expectation value
of $\phi$ at $T_c$ in the broken phase
computed from it are $T_c = 173.3$ GeV and $v(T_c)= 81$ GeV
($v(T_c)/T_c = 0.47$).
\section{The lattice action}
Going over to a 2$\times$2 matrix representation $\Phi=
(\phi_0+i\sigma_i\phi_i)/\sqrt2$  of the doublet scalar field,
discretising and scaling the continuum fields by
\be
igaA_0\to A_0,\qquad\qquad \Phi\to \sqrt{{T\over
a}{\beta_H\over2}}\Phi,
\la{rescaling}
\ee
the lattice action corresponding to the continuum action
in eq.~(\ref{3daction}) becomes, in standard notation,
\begin{eqnarray}
&&S= \beta_G \sum_x \sum_{i<j}(1-\fr12 \tr P_{ij}) +\nonumber \\
&&+ \fr12\beta_G  \sum_x \sum_i[\tr
A_0(\bfx)U_i^{-1}(\bfx)A_0(\bfx+i)
U_i(\bfx) - \tr A_0^2(\bfx)]+
\nonumber \\
&& + \sum_x \biggl\{10\Sigma - {5\over3}
{4\over g^2\beta_G} \biggr\} \fr12 \tr A_0^2(\bfx) + \nonumber \\
&& + \sum_x {g^2\beta_G\over 3\pi^2} \left(\fr{17}{16}\right)
\left( \fr12 \tr A_0^2(\bfx) \right)^2 + \la{latticeaction}\\
&& + \beta_H \sum_x \sum_i\biggl[
\fr12\tr\Phi^\dagger(\bfx)\Phi(\bfx)
- \fr12\tr\Phi^\dagger(\bfx)U_i(\bfx)\Phi(\bfx+i) \biggr]+\nonumber
\\
&&+ \sum_x \biggl[ (1-2\beta_R - 3\beta_H)
\fr12\tr\Phi^\dagger(\bfx)\Phi(\bfx) + \beta_R
 \biggl( \fr12\tr\Phi^\dagger(\bfx)\Phi(\bfx) \biggr)^2 \biggr]
+\nonumber \\
&& - \fr12\beta_H \sum_x
\biggl[\fr12\tr A_0^2(\bfx) \fr12\tr\Phi^\dagger(\bfx)\Phi(\bfx)
\biggr].\nonumber
\end{eqnarray}
where $\Sigma=0.252731$.

All the three lattice  coupling constants are
given in terms of $g, T$ and $m_H$ by the following
equations, which directly follow from the discretisation procedure:
\begin{eqnarray}
&&\beta_G = {4\over g^2}{1\over Ta}, \label{betag}\\
&&\beta_R={1\over4}\lambda Ta \beta_H^2 =
{m_H^2\over 8m_W^2}{\beta_H^2\over\beta_G}. \la{betar}
\label{betas}
\end{eqnarray}
The tree relation between the mass parameter $m_3$ and the lattice
variables is given by
\be
m_3^2={2(1-2\beta_R-3\beta_H) \over \beta_H a^2}.
\ee
 The 1-loop counterterm ($\sim 1/a$) removing divergences from the
Higgs mass has been determined in \cite{kari}, and the 2-loop
counterterm $\sim \log(a)$ has been estimated in \cite{fkrs}.
For present lattices the linear counterterm, included in \cite{kari},
is clearly the dominant one, but the accuracy of the calculations is
already such that the effects of the logarithmic counterterm are
clearly seen. No other counterterms exist since the 3d
theory is super-renormalizable.

The knowledge of these counterterms allows one to relate lattice
parameters to the temperature:
\begin{eqnarray}
&&{m_H^2 \over4T^2}=\biggl({g^2\beta_G\over4}\biggr)^2
\biggl[ 3-{1\over\beta_H} + {m_H^2\over4m_W^2} {\beta_H\over\beta_G}
-{9\over2\beta_G}\left(1+{m_H^2\over 3m_W^2}\right)\Sigma -
\nonumber\\
&&-\fr12 \biggl({9\over 4\pi\beta_G}\biggr)^2
\biggl(1+{2m_H^2\over9m_W^2}-{m_H^4\over27m_W^4}\biggr)
\biggl(\log{g^2\beta_G\over2}+\eta\biggr)\biggr]\nonumber\\
&&+{g^2\over2}\biggl[
\fr3{16}+  {m_H^2\over16m_W^2} + {g^2\over16\pi^2}\biggl({149\over96}
+{3m_H^2\over 32m_W^2}\biggr)\biggr].\la{betahT}
\end{eqnarray}
Because of the super-renormalizability of the 3d theory, this
relation is
exact in the continuum limit. On the 1-loop level the analogous
relation has been derived in \cite{kari}.
The determination of the ($m_H/m_W$ dependent) constant $\eta$
requires a computation of the 2-loop effective potential of the
theory
by the lattice perturbation theory and a comparison of the result
with
the
corresponding expression in the ${\overline{{\rm MS}}}$ scheme. Due
to
the
complexity of this computation, it is not attempted here. Instead, we
determine the constant $\eta$ by Monte--Carlo methods.
\section{Lattice results}
The continuum limit corresponds to $\beta_G\to\infty$,
$N/\beta_G \rightarrow \infty$, with $N$ being the lattice size.  Too
large a
$\beta_G$ makes the system hard to simulate and we have used
$\beta_G=12,20,32$. Our computer resources have permitted us to
perform runs on lattices of sizes $8^3$ ... $32^3$. The confining
property of the symmetric phase must not be lost, and this requires
\be
\beta_G<1.468N,
\label{conf}
\ee
at least for pure 3d SU(2) gauge theory \cite{teper}.
The use of different values of $\beta_G$ and $N$ is important to
study the scaling and consistency of the results.

Simulations have also been carried out for the Higgs masses $m_H=
35,~60,~
70,~80$ and $90$ GeV. We will discuss here in some detail only those
simulations with $80$ GeV Higgs (other Higgs masses are discussed in
\cite{fkrs2}).

The first question is the very existence of the first-order phase
transition. To answer it, we searched for a two-peak signal in the
distributions of the different order parameters (such as length of
the
Higgs field $R^2$, the average of the link operator
$L=\tr V^{\dagger}(x)U_i(x)V(x+i)$, where $\Phi = RV$, etc.) A
typical picture at
$\beta_H=0.347710$, $\beta_G=12$ and $V=24^3$ is shown in Fig. 1.
The two-peak structure is clearly seen. At the same time, due to the
fact that the lattice volume is finite, the width of any of the
peaks is comparable with the distance between them. For comparison,
we present in Fig. 2 the distribution of $R^2$ for a Higgs mass
$m_H=35$ GeV, where finite-size effects are less important.

The determination of the critical value of $\beta_H^c$ in the
continuum limit requires the analysis of finite-size effects. In
principle, a number of different methods can be used for its
determination:\\
(i) Equal-area signal. The value of $\beta_H^c$ is determined
as the value of $\beta_H$ for which the {\em areas} \cite{book}
under the two peaks in the distribution of some order
parameter (actually, it
does not matter which order parameter is chosen) are equal. \\
(ii) One looks for a maximum of the heat capacity (action
susceptibility), considered as a function of $\beta_H$
\cite{book,borgs}.\\
(iii) The same as in (ii), but for $L$-susceptibility
\cite{borgs,ferrenberg}.

All three methods must give the same results in the
continuum limit.
For finite lattices and for finite $\beta_G$ the results are
different, and the continuum limit should be found by
extrapolation.

The first method works well for small Higgs masses, such as
$35$ GeV, since the two peaks are well separated, but it cannot be
applied with good accuracy to the study of the phase transition with
heavy enough Higgses, at least for the lattice sizes we used. So, we
used (ii) and (iii) to determine $\beta_H^c$. To find the maximum
values,
we used the Ferrenberg-Swendsen multihistogram method
\cite{ferrenberg}, which
yields Monte Carlo observables as continuous functions in $\beta_H$
around the actual $\beta_H$ -values used in the simulations. The
error analysis was performed with the jackknife method.

The results of the determination of the critical $\beta_H^c$ are
shown in Fig. 3 ($\beta_G=12$) and Fig. 4 ($\beta_G=20$) as a
function
of $1/N^3$, with $N= 8,~12,~16,~20,~24$ and $32$ (without the point
$N=20$ for $\beta_G= 20$). One can see that both methods give
consistent results for sufficiently large volumes of the lattice.

The extrapolation of these results to infinite volume requires
some care. For $\beta_G=12$ eq. (\ref{conf}) gives $N>8$. The
$\beta_G=12$ data for $N \geq 12$ can be fitted very well by a linear
function $\beta_H^c(V) = \beta_H^c(\infty) + c/V$, giving
$\beta_H^c(\infty)=
0.347703(10)$. The power low fit $\beta_H^c(V) = \beta_H^c(\infty) +
c V^n$ for all lattice volumes gives a consistent result:
$\beta_H^c(\infty)=
0.347698(12)$ with the power $n=0.9 \pm 0.1$.

For $\beta_G=20$ finite-size effects are more important. From eq.
(\ref{conf}) one gets for $\beta_G=20$ $N>14$, so that one to two
points at the right  of Fig. 4 cannot be trusted. Then, the linear
fit with the use of the four largest volumes gives
$\beta_H^c(\infty)=0.341721(9)$; with the three  largest volumes,
$\beta_H^c(\infty)=0.341710(11)$;with the three  largest volumes,
$\beta_H^c(\infty)=0.341668(21)$.

Our final aim is the determination of the critical temperature. To
fix it, we must know the parameter $\eta$ appearing in eq.
(\ref{betahT}). Since there are no analytic computations of this
parameter with
the help of lattice perturbation theory  (even if they are possible),
we determine it by comparison of the results of lattice simulations
with perturbation theory deep in the broken phase, where it
should work well. We choose the
order parameter $R^2$ for this purpose. In order to relate it to
continuum physics one has to subtract from it linear ($\sim 1/a$) and
logarithmic ($\sim \log (a)$) divergent terms. It can be shown that
$\langle R^2\rangle$ is related to the continuum effective potential
through
\be
\beta_H \langle R^2\rangle= \left[4\Sigma +\frac{3}{2\pi^2\beta_G}
\left(\log\frac{3g_3^2\beta_G}{2\mu}+\bar{\eta}\right)\right]+
\frac{8}{g_3^2\beta_G}\frac{dV_{eff}(v(T))}{dm_3^2},
\label{voverT}
\ee
where the expression in square brackets contains terms that are
divergent in the continuum limit, $\mu$ is an arbitrary normalization
point (the total expression is $\mu$-independent due to the $\mu$
dependence of the effective potential $V_{eff}$), $\bar{\eta}$ is
unknown constant. The derivative of the effective potential is taken
at $\phi$ in the minimum of the broken phase. In the leading
approximation, $\frac{dV_{eff}}{dm_3^2}|_{v(T)}= \fr12 v(T)^2$. The
two-loop counter-term contribution to expression (\ref{voverT}) is
numerically suppressed, so that uncertainties in the determination of
$\eta$ coming from an absence of information about $\bar{\eta}$ are
small provided $\bar{\eta}$ is not too large (say, $\bar{\eta}<2$).

The evolution of $R^2$ distributions in the broken phase with
$\beta_H$ is shown in Fig. 5. In Fig. 6 we show $\langle R^2\rangle$
as a function of $\beta_H$ both as given by lattice data and by
2-loop perturbative computation with parameter $\eta=0.54$
($\beta_G=12$) and $\eta=0.12$ ($\beta_G=20$). The $\beta_G =12$
lattice data corresponds to the variation of $v(T)/T$ from $1.3$ to
$2$, and $\beta_G=20$ data to $v(T)/T = 1.7$--$3.9$. One can see that
the use of 2-loop effective potential for the determination of the
vacuum expectation value for the Higgs field in the {\em broken
phase} at {\em given temperature} works quite well (numerically
$\frac{\delta v(T)}{v(T)}< 1$--$2\%$). This means that the
dimensionless expansion parameter $\sim g^2 T/m_W(T)$ is sufficiently small
at these temperatures. In other words, non-perturbative effects and
higher order terms can be essential numerically only for $v(T)/T <1$.

At the same time, the systematic difference between the lattice data and
2-loop continuum predictions is clearly visible (it is, though, very small,
less than $1\%$ in $\langle R^2\rangle$). In Fig. 7 we show the the plot of
$R^2_{latt}-R^2_{theor}$ as a function of $\beta_H$ and with the same $\eta$
choices as in Fig. 6 (the best fits). We suspect that the main source of the
difference comes from finite size and finite spacing effects, which are
difficult to compute analytically. As a rough estimate of parameter $\eta$ we
take  $\eta = 0.3 \pm 0.5$, the central value being the average of $\beta_G=12$
and $\beta_G=20$ best fit values, while the error estimate is the difference
between them.

With the value of $\eta$ determined, the observed values of
$\beta_H^c$ can
be converted to results for the critical temperature of the
electroweak
phase transition using eq. (\ref{betahT}). From the $\beta_G=12$
data we get $T_c = 162.1 \pm 2.6$ GeV and from the $\beta_G=20$ data
quite consistent values
$160.3 \pm 2.6,~160.9 \pm 2.8$ and $163.4 \pm 3.0$ GeV, depending on
the extrapolation to
infinite volume (using four, three, two largest volumes,
respectively). One observes that the $T_c$
determined by lattice methods
is clearly smaller than the value $T_c$ = 173.3 GeV determined by
perturbative methods.

Since we do not have a clear peak separation at the critical
temperature (implying that the system contains, at this
temperature, a large fraction of interface configurations
interpolating between the broken and symmetric phases) we cannot
unambiguously
extract from the lattice data at $T_c$ the quantity $\langle
R^2\rangle$ relevant for the broken phase only. However, if we take
the position of the peak to the right of Fig. 1 as an
estimate of $\langle
R^2\rangle$ in the broken phase, then we see
from eq. (\ref{voverT}) that $v(T_c)/T_c = 0.73 \pm 0.04$
(the error estimate is based on the width of the peak). This value
may be compared with the results computed in
optimized 2-loop perturbation theory: $v(T_c)/T_c = 0.47$ and
$v(T)/T = 0.81$ for, say, $T=163.6$ GeV (which is taken as a "true" critical
temperature). We conclude that the true jump of the order
parameter (0.73) is clearly larger than
that (0.47) coming from perturbation theory. At the same time,
the perturbative prediction of $v(T)$ at $T=163.6$ GeV
is rather close to the true value (compare $0.73$ and $0.81$).
The difference may be due to the fact that the
unambiguous extraction of $v(T)/T$ from the distribution of
$R^2$ is not possible and that finite size effects are not small. Another
possibility is that at $v(T)/T \sim
0.7$ deviations (perturbative or non-perturbative) from 2-loop
potential predictions are as large as 10\%.
\section{Discussion}
The lattice results indicate that the phase transition is more
strongly
first order than could be expected from perturbation theory. At the
same time, a good convergence of the perturbation theory in the
broken phase allows us to relate these deviations with
the properties of the symmetric phase. At present we cannot extract
from available lattice data the parameters relevant for
cosmological applications. In particular, the region of metastability
of the symmetric phase, together with the bubble nucleation
temperature
$T^*$ cannot be determined. The direct lattice determination of $T^*$
is hardly possible at all, so that some indirect methods are
necessary. One of them is related to the measurement of the
interface tension at the critical temperature, which allows one to
compute the bubble nucleation rate at least in the vicinity of the
critical temperature. Another obvious option is to study details of
the phase transition for the heavy Higgs boson, since we are
approaching the non-perturbative region from the side of
the broken phase,
increasing $m_H$. We plan to return to these questions in future
work.

As has been discussed in \cite{ms}, non-perturbative effects are
likely to decrease the value of the potential in the vicinity
of the origin. It seems that our lattice results are in a perfect
qualitative agreement with this picture. Indeed, consider the
plot of the 1-loop and 2-loop renormalization group improved
effective potentials at the "true" critical temperature $T_c^{true}
\simeq 163.6$
shown in Fig. 8 (note that the deviation of the 1-loop result
from the 2-loop
one is negligibly small thanks to the optimization procedure,
\cite{fkrs}).
Clearly, these potentials are completely wrong at small fields
$\phi$, since we know that {\em we are} at the critical temperature.
Therefore, the effective potential must have a contribution making
the broken and symmetric phases degenerate. According to Fig. 8
this contribution at $\phi =0$ equals
\be
V_\rmi{pert}(v(T_c)) - V_\rmi{pert}(0)\approx
0.03 g_3^6 \approx 60 (\alpha_W T)^3.
\ee
A possible
non-perturbative contribution is shown in this picture by a dashed
line
(in notations of ref. \cite{ms}, we have taken $A_F=0.36$ in the
non-perturbative part of the potential).
If the specific model for the description of non-perturbative effects
considered in \cite{ms} is correct, then with this value of the
non-perturbative energy shift electroweak baryogenesis may be
possible up to a Higgs mass of about $100$ GeV. The exact
determination
of this bound, however, requires a much better understanding of the
symmetric phase of the electroweak theory at high temperatures.

\newpage

\begin{figure}
\psfig {file=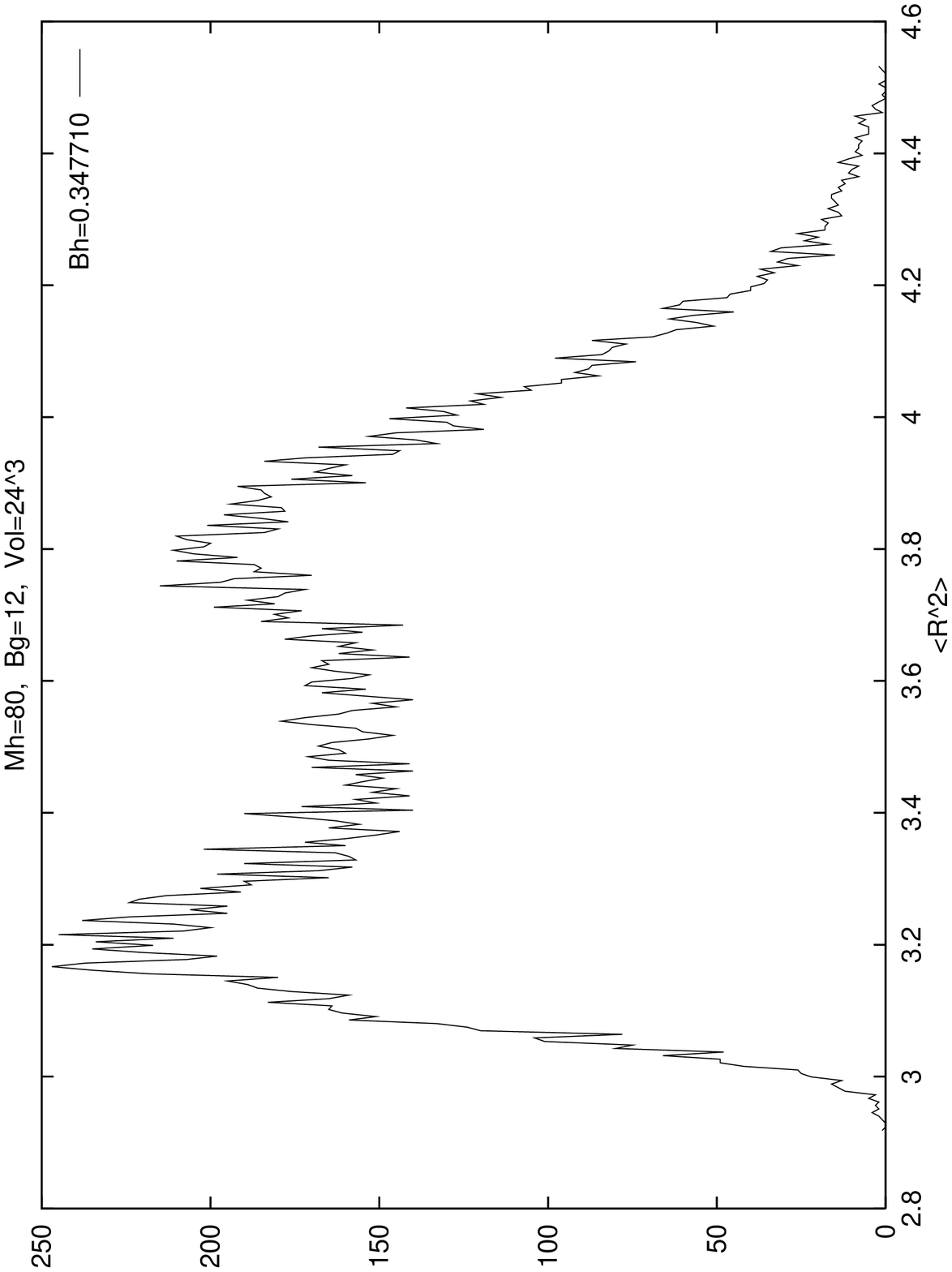,width=5in}
\caption{Distribution of the order parameter
$R^2=\fr12\tr\Phi^\dagger\Phi$ at the critical temperature.}
\end{figure}

\begin{figure}
\psfig {file=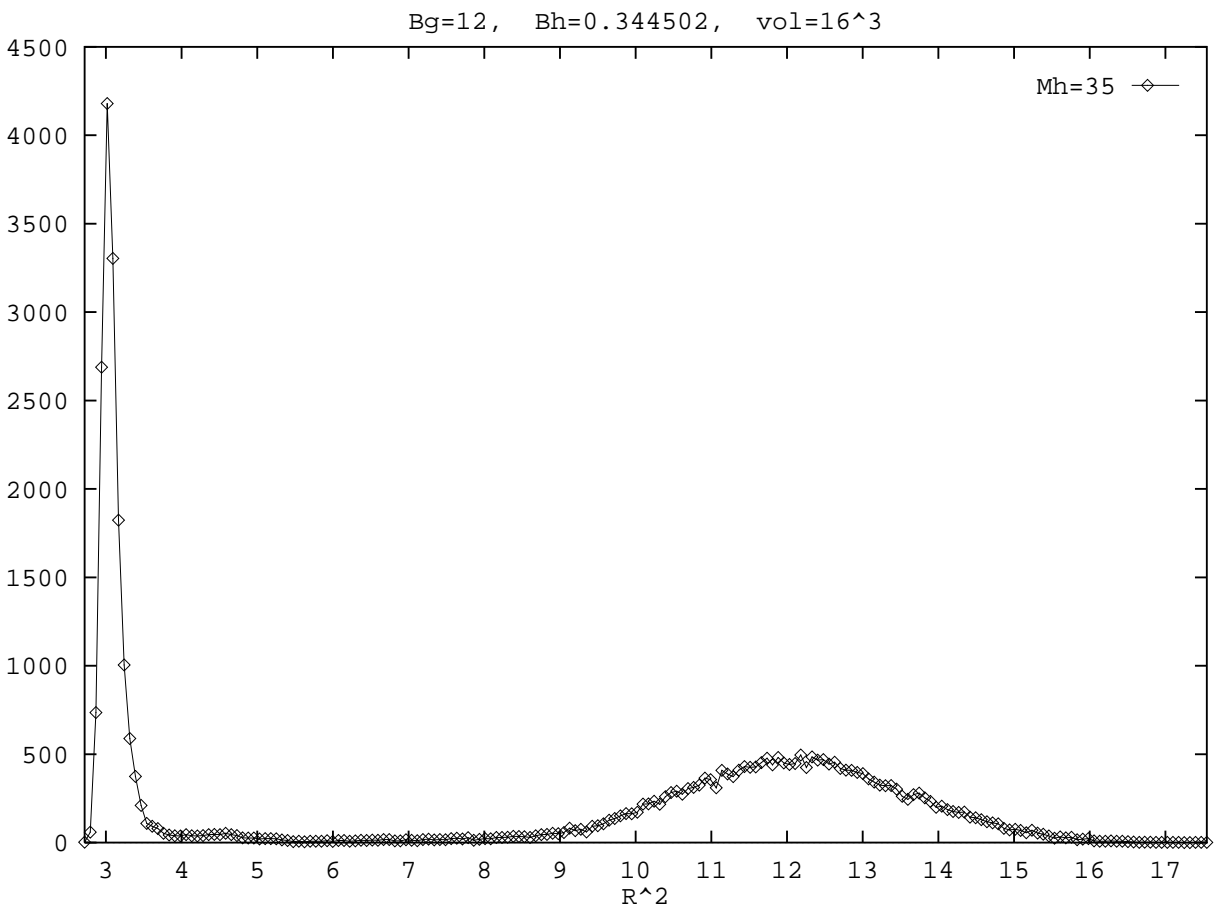,width=5in}
\caption{Distribution of the order parameter $R^2$ at the critical
temperature, with $m_H=35$ GeV.}
\end{figure}

\begin{figure}
\psfig {file=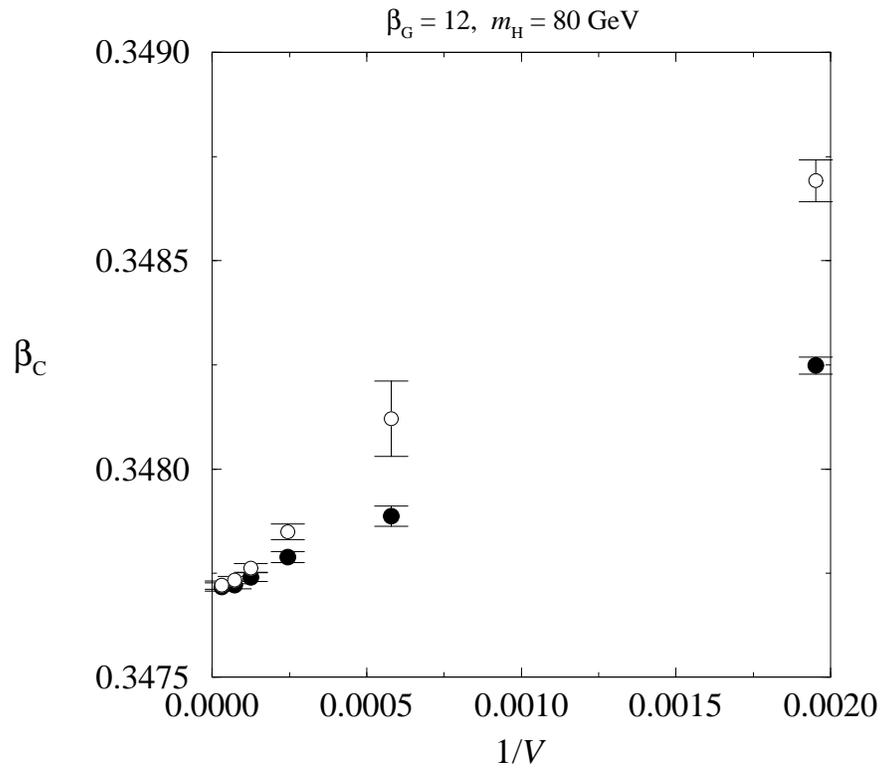,width=5in}
\caption{The critical values of $\beta_H$ for $m_H=80$ GeV and
$\beta_G=12$ for different lattice sizes. The full points correspond
to
the maximum of $L$-susceptibility, and empty points to the maximum
of the heat capacity.}
\end{figure}

\begin{figure}
\psfig {file=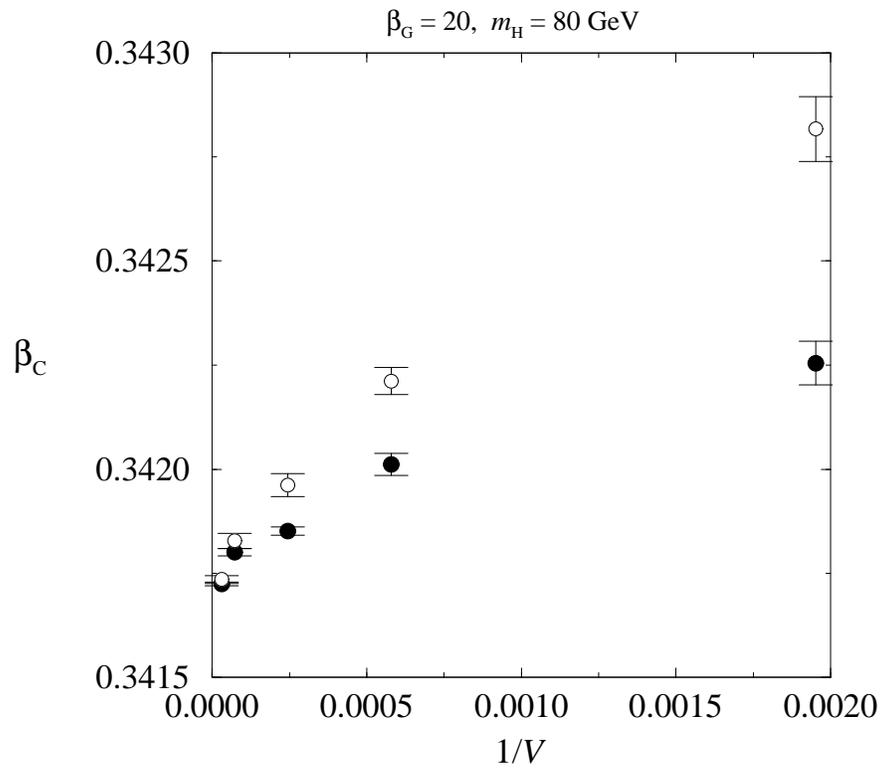,width=5in}
\caption{The same as Fig. 3 but for $\beta_G=20$ }
\end{figure}

\begin{figure}
\psfig {file=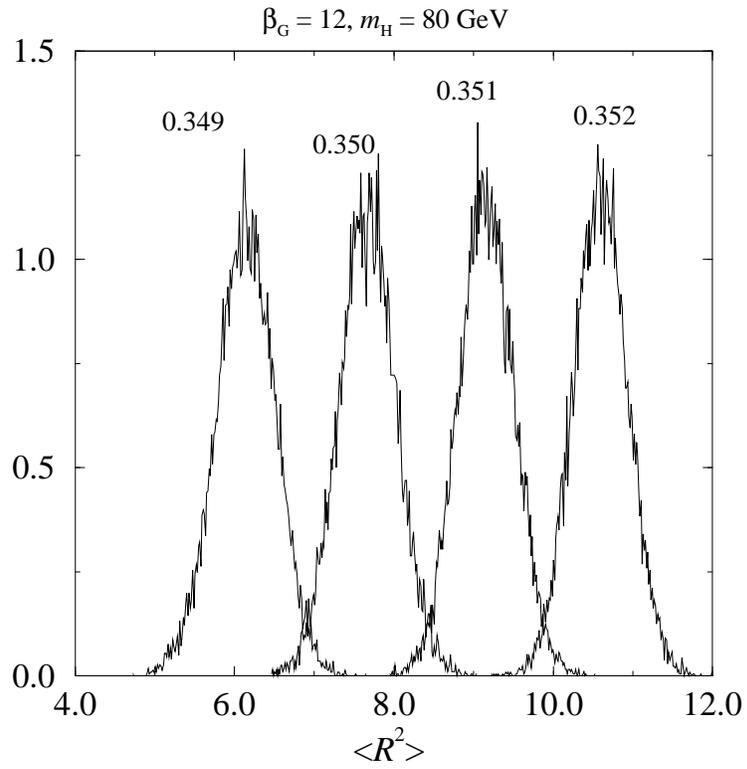,width=5in}
\caption{Evolution of the distribution of $R^2$ in the broken phase
with
$\beta_H$ for $\beta_G = 12$, with $m_H = 80$ GeV.}
\end{figure}

\begin{figure}
\psfig {file=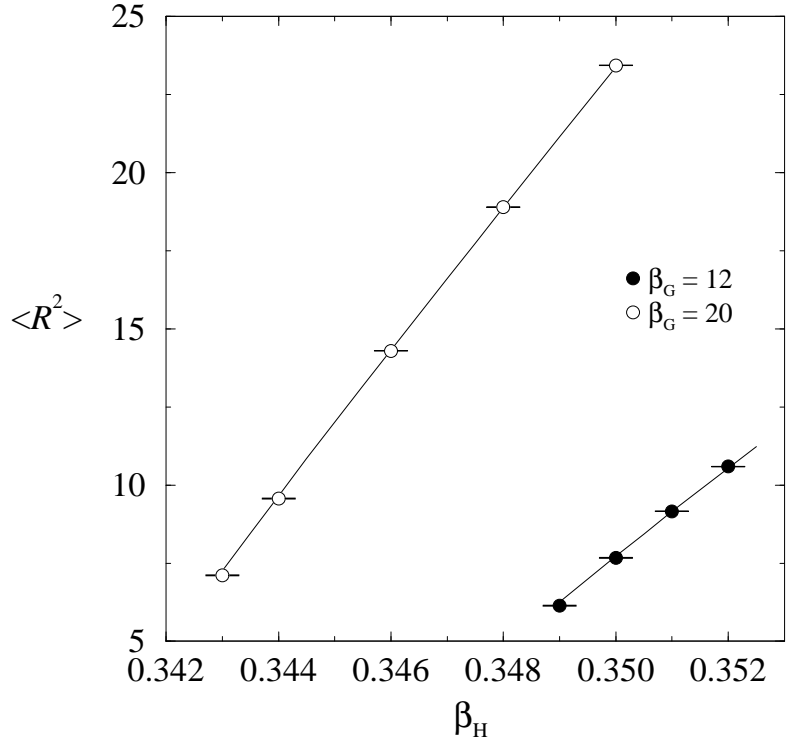,width=5in}
\caption{$\langle R^2\rangle$ determined on the lattice (points) and
by perturbation theory (solid lines). The upper curve
corresponds to $\eta=0.12$, $\beta_G = 20$ and the lower one to $\eta=0.54$,
$\beta_G=12$. }
\end{figure}

\begin{figure}
\psfig {file=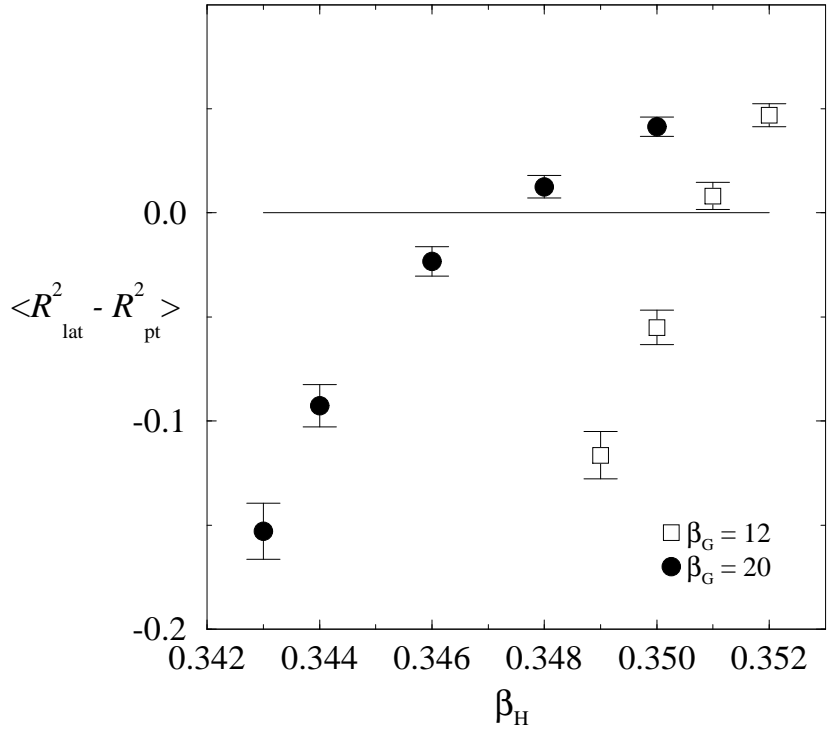,width=5in}
\caption{The difference between $\langle R^2\rangle$ determined on the lattice
and  $\langle R^2\rangle$ found with the use of  perturbation theory for
$\eta=0.12$, $\beta_G = 20$ and  $\eta=0.54$, $\beta_G=12$. }
\end{figure}

\begin{figure}
\psfig {file=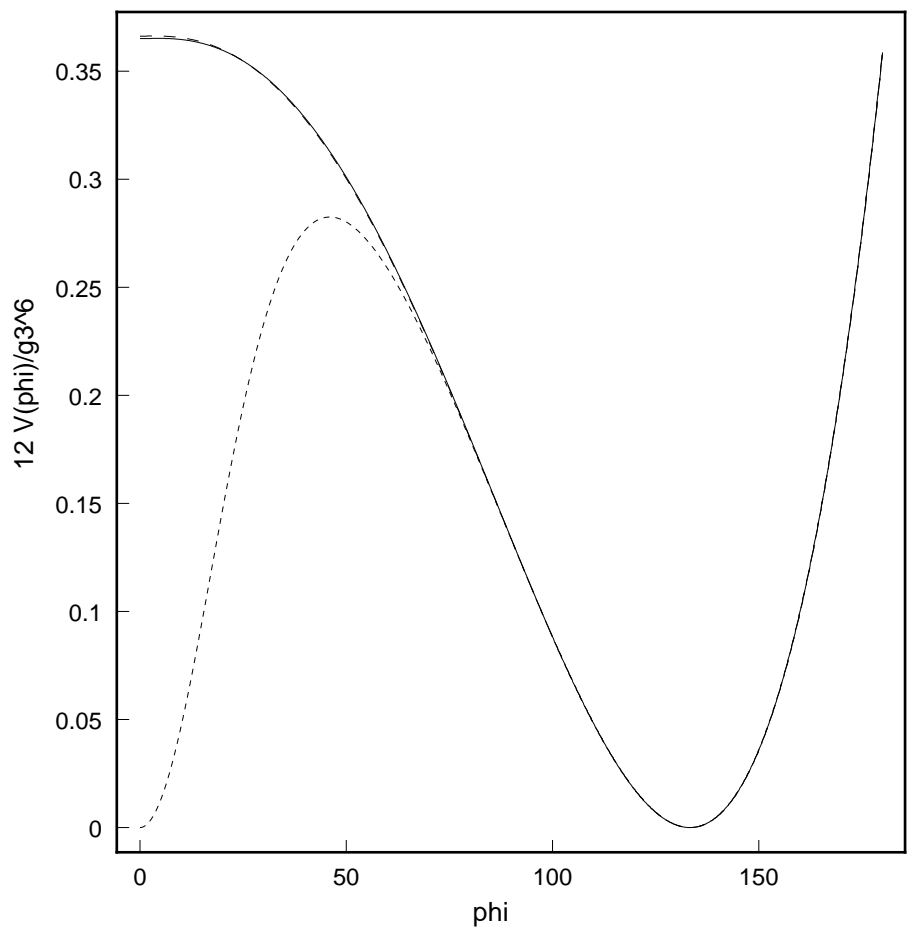,width=5in}
\caption{The renormalization group improved effective potential.
Solid line  -  2-loop order, dashed line--1-loop order, short-dashed
line -- possible non-perturbative contribution. The $x$-axis is the
4d scalar field in GeV, the $y$-axis is the dimensionless
effective potential $12 g_3^{-6}V_{eff}(\phi)$.}
\end{figure}
\end{document}